\RequirePackage[2020-02-02]{latexrelease}
\documentclass[nofootinbib,aps,letterpaper,preprintnumbers,superscriptaddress,showkeys,11pt]{revtex4}
\usepackage{graphicx,epsfig}
\usepackage{amsmath}
\usepackage{amsfonts}
\usepackage{amssymb}
\usepackage{amsmath, amssymb, graphicx}
\usepackage{epsf}
\usepackage{epstopdf}
\usepackage {amssymb}
\usepackage{array}
\usepackage[colorlinks]{hyperref}
\usepackage[usenames,dvipsnames]{color}
\definecolor{beamer@PRD}{RGB}{46,48,146}
\definecolor{oxfordblue}{rgb}{0.0, 0.13, 0.28}
\definecolor{burgundy}{rgb}{0.5, 0.0, 0.13}
\definecolor{darkolivegreen}{rgb}{0.33, 0.42, 0.18}
\definecolor{bluer}{rgb}{0.00,0.50,0.75}
\definecolor{richcarmine}{rgb}{0.84, 0.0, 0.25}
\definecolor{darkgreen}{rgb}{0.00,0.50,0.25}

\hypersetup{
     breaklinks=true,
    pdfstartview={FitH},    
    colorlinks=true,       
    linkcolor=magenta,          
    citecolor=cyan,        
    filecolor=magenta,      
    urlcolor=richcarmine,           
    anchorcolor=red,      
    linktocpage=true
}

\begin{document}
\date{\today}
\newcommand\be{\begin{equation}}
\newcommand\ee{\end{equation}}
\newcommand\bea{\begin{eqnarray}}
\newcommand\eea{\end{eqnarray}}
\newcommand\bseq{\begin{subequations}} 
\newcommand\eseq{\end{subequations}}
\newcommand\bcas{\begin{cases}}
\newcommand\ecas{\end{cases}}

\newcommand\tcr{\textcolor{red}}

\newcommand{\p}{\partial}
\newcommand{\f}{\frac}

\title{A Survey of Strong Cosmic Censorship Conjecture Beyond Einstein's Gravity}

\author {\textbf{Mohsen Khodadi}}
\email{m.khodadi@ipm.ir}
\affiliation{School of Astronomy, Institute for Research in Fundamental Sciences (IPM),\\
	P. O. Box 19395-5531, Tehran, Iran}
\author{\textbf{Javad T. Firouzjaee}}
\email{firouzjaee@kntu.ac.ir} 
\affiliation{Department of Physics, K. N. Toosi University of Technology, \\
	P. O. Box 15875-4416, Tehran, Iran}
\affiliation{ School of Physics, Institute for Research in Fundamental Sciences (IPM), \\
	P. O. Box 19395-5531, Tehran, Iran } 

\begin{abstract}
The strong cosmic censorship conjecture (SCCC) proposed by Penrose states that the presence of the inner Cauchy horizon ($\mathcal{CH}$)
in the black hole solutions does not threaten the deterministic nature of general relativity since it is actually unstable
against the remnant perturbation fields that fall into the black hole.  Although this conjecture is well-established for the asymptotically flat spacetimes in general relativity,  it is challenged upon taking a small positive cosmological constant as
required by observational data. This challenge can be fixed by finding even one type of quasi-normal mode whose decay rate is slow enough insofar as the remnant of perturbation is able to destroy the $\mathcal{CH}$. By aiming to investigate the role of the additional parameter induced by modified gravity in this scenario,  we consider the charged-de Sitter black hole solutions of Einstein-Maxwell theory extended with $f(R)$ and Energy-Momentum Squared Gravities. Employing the numerical analysis of the deterministic criterion, we show that the presence of the model parameter of theories at hand within a given range controls the decay rate of the photon sphere quasi-normal modes (in the eikonal limit) to rescue SCCC.
 
 \end{abstract}
\maketitle

\section{Introduction}

The ability to predict the future of the physical phenomena using initial data is one of the salient features of classical theories such as Newtonian mechanics and general relativity (GR).
Given the importance of initial data, GR is supported by a fundamental theorem \cite{ChoquetBruhat:1969cb} that asserts there exists a so-called maximal Cauchy development as the largest spacetime determined uniquely by the initial data. However, there is always room for the question of whether could the maximal Cauchy development be a subset of a larger spacetime so that the larger spacetime could not be predicted from the initial data?
If yes, we will face the defeat of determinism because we no longer would be able to use the initial data to predict the state of spacetime in the future. In order to preserve the deterministic nature of GR, Roger Penrose formulated a strong cosmic censorship conjecture (SCCC)
\footnote{It is important to stress two points. First, despite there is currently no solid mathematical proof to support  SCCC in curved background \cite{Hod:2019zoa}, however, there is a newly provided proof based on Bekenstein's generalized second law of thermodynamics \cite{Hod:2020ktb}. Second, Penrose also proposed a weak cosmic censorship conjecture logically unrelated to SCCC, which asserts that when the gravitational collapse of a star causes a spacetime singularity to form, this singularity is beyond the reach of any distant observer by hiding inside the black hole \cite{Penrose:1969pc}.} \cite{Penrose1,Hawking} that prohibits such a possibility.
Broadly speaking, the SCCC states that all the physically reasonable solutions of Einstein's equations are globally hyperbolic, meaning that they enjoy the properties of uniqueness, continuity, and predictability \cite{Book}.
However, for the Reissner-Nordstr\"{o}m (RN) and Kerr black hole solutions of Einstein's equation, we encounter a challenging situation.
More exactly, in these solutions, the maximal Cauchy development does not cover the entire spacetime, but is limited to a boundary so-called the Cauchy horizon ($\mathcal{CH}$) inside the black hole. The challenge arises from the fact that beyond $\mathcal{CH}$, the spacetime may smoothly be extended in many different ways, all satisfying Einstein's equation \cite{Chandra}.
So due to losing causal connection after safely crossing the observer from the $\mathcal{CH}$, the initial data and Einstein's equation no longer can predict the observer's future, indicating an open violation of SCCC. However, in the light of leading works \cite{Simpson:1973ua,Poisson:1990eh,Dafermos:2003wr,Dafermos:2012np}, it turns out that in RN black hole solution does not occur a violation of SCCC since the $\mathcal{CH}$ is unstable.
Namely, in case of perturbating the initial data so that it slightly deviates from the spherical symmetry, the resulting perturbed spacetime does not expose a $\mathcal{CH}$ but instead ends at a singularity. The perturbating on any asymptotically flat black hole, in essence, results in arise of a phenomenon so-called \emph{``mass inflation singularity''} on $\mathcal{CH}$ that is strong enough to breakdown the field equations and subsequently save SCCC. More technical, the mass inflation singularity comes from the presence of time-dependent remnant fields in the external spacetime regions, which decay asymptotically slow enough in the form of power-law. This renders a mechanism with an exponential blue-shift amplification of matter and radiation fields inside black holes, which fixes the pathological aspect of $\mathcal{CH}$ via turning it into spacetime singularities \cite{Hod:1997fy,Hod:1998ra,Hod:1999rx}. Note that it is right just for asymptotically flat black holes.
To comprehensive understanding this phenomenon from the perspective of some alternative views, it is good to refer to papers such as \cite{Ori:1991zz,Christodoulou:2008nj,Ringstrom:2015jza}.

However, Einstein's equation includes the cosmological constant $\Lambda$, which can explain the universe's acceleration. Indeed
in light of the observational data, it is well-known that the value of the cosmological constant should be positive and very small, as well. By taking into account a positive cosmological constant in the background of spacetime, it has been demonstrated that the fate of stability/instability of $\mathcal{CH}$ is dependent on a delicate competition between two opposite physical mechanisms:
\begin{itemize}
  \item It is well-known that astrophysical black holes in a de Sitter (dS) universe stay safe against perturbations due to the exponentially decaying   ($\sim e^{-g t}$) of the neutral matter fields propagating in the external spacetime regions \cite{Brady:1996za,Chambers:1997ef}.
  Compared to its counterpart in a universe free of a cosmological constant, which obeys the form of power-law decaying, it decays faster
  \cite{Gundlach:1993tp,Gundlach:1993tn}. The positive constant $g$ here is the rate of decay and encoded by the imaginary part of the fundamental or longest-lived quasi-normal modes (QNMs) attributed to a composed black hole-field system i.e., $-\{Im ~(\omega_n)\}_{min}$.
  
  \item The external remnant fields are subject to an exponential boost $\sim e^{\kappa_c \upsilon}$ as they fall into the black
  holes and propagate parallel to the $\mathcal{CH}$ \cite{Brady:1996za}. Here $\upsilon$ and $\kappa_c$ are respectively a standard advanced null coordinate and the surface gravity of the $\mathcal{CH}$. The latter, in essence, represents the characteristic rate of the blue shift mechanism.
  \end{itemize}
As a result, the final status of RN-dS and Kerr $\mathcal{CH}$ stability is determined due to competition between the decay rate of fundamental damped QNMs and the rate of blue shift arising from the surface gravity $\kappa_c$ through the following dimensionless factor \cite{Hintz:2016gwb,Dafermos:2017dbw,Cardoso:2017soq}
\begin{align}\label{beta}
\beta\equiv -\frac{\{Im ~(\omega_n)\}_{min}}{\kappa_c}~.
\end{align}
In case of extracting a finite range of the black hole parameter spaces in which for all of possible QNMs the inequality $\beta\geq1/2$ is satisfied, then the relevant black hole spacetime physically is allowed to an extension to beyond its $\mathcal{CH}$, signaling the open failure of determinism. Hence it would not be surprising if we call $\beta$ a \emph{deterministic factor}  because one can imagine it as a sufficient criterion to survey the validity of SCCC \footnote{Notice that the criterion $\beta<1/2$ is tighter than its old version i.e., $\beta<1$ in order to preserve SCCC. More clearly, it is argued that the inequality $\beta<1$ is merely related to the blow-up of curvature invariants which does not matter much in the sense that it does not guarantee the breakdown of field equations and maybe still find some safe weak solutions on the $\mathcal{CH}$ \cite{Klainerman:2012wt,Klainerman:2012wy}.}. It is important to pay attention to two points. 
First, it is customary to calculate the $\beta$-criterion for three well-known QNMs: photon sphere (PS), de Sitter (dS), and the near extremal (NE) \cite{Cardoso:2017soq}. Second, if inequality $\beta\geq1/2$ is not met even by one of these QNMs, the SCCC still remains valid.

From a theoretical point of view, this is very well-motivated via a robust and quantitative manner to  study the validity/invalidity of the Penrose SCCC in dS black hole backgrounds subjects to a variety of perturbations, see \cite{Dias:2018ynt}-\cite{Rahman:2019uwf}.
Although the outcomes of these studies contain worrying messages about the Penrose SCCC violation within just RN-dS black holes, such a concern also may exist for the astrophysical black holes (Kerr) \cite{Casals:2020uxa}.

Despite all the brilliant achievements of GR, it is not free of shortcomings (particularly at infrared and ultraviolet scales), and it is merely a solid effective theory for modeling gravitational physics, which should be completed by some extensions \cite{Capozziello:2011et}. This has led to people do not limit their studies concerning the validity of Penrose SCCC just for the GR framework rather in recent years we are seeing a variety of works in the context of some extended models of gravity \cite{Rahman:2018oso}-\cite{Singha:2022bvr}. Given extracting some counterexamples in favor of violation of SCCC within GR, this can be seen as a natural attempt to save the SCCC via going beyond the GR framework. However, the results reported from the above-listed references not only are not encouraging but in some cases exacerbate the status of SCCC violation, too. It would be interesting to mention that the attempt to rescue SCCC via introducing a charged scalar perturbation around the RN-dS metric also has failed \cite{Dias:2018ufh}.


To that end, in this paper, we plan to extend the study of the validity of Penrose SCCC for RN-dS black holes within further modified Einstein-Maxwell models with the hope of restoring credibility to SCCC.
First, we investigate the tail of effective potential corresponding to these two models at long distances. We then consider whether, in the extended charged spherically symmetric backgrounds, the decay of scalar wave-tail at the late time still occurs with a rate faster than case $\Lambda=0$ or not?
To investigate the validity of Penrose SCCC, by utilizing the geometric-optics technique we focus on PS-QNMs. We will show that in the light of extension made by the underlying gravity models in a system composed of the black hole and scalar field, it is possible to extract a parameter space in which condition $\beta\geq1/2$ is not satisfied.
As a result, circumventing the mentioned inequality by only one type of QNMs (here PS) is sufficient to restore credibility to SCCC. Specifically, the gravitational extensions under our consideration in this paper come from: $f(R)$-Gravity (Sec. \ref{sec.FR}), and  Energy-Momentum Squared Gravity (Sec. \ref{sec.EMSG}). Finally, we will end this paper in Sec. \ref{con}.
Throughout this paper, for simplicity, we work with the units $c=G_N=\hbar=1$ and $\kappa\equiv 8\pi G_N=1$.

\section{Connection between deterministic inequality and validation of SCCC } \label{sec.II}

To show how the dimensionless factor $\beta$ is related to the SCCC, we consider a massless scalar field living on any $d$-dimensional static and spherically symmetric spacetime which satisfies Klein-Gordon's equation $\square \Psi=0$.
Given that the spacetime enjoys
the existence of angular and timelike Killing vectors so the scalar field can be decomposed as
$\Psi(t,r,\Omega)=e^{-i\omega t}R(r)Y(\Omega)$, where respectively $Y(\Omega)$ and $R(r)$ represent
the spherical harmonics associated with the $(d-2)$ dimensional sphere and a radial function satisfying
a second order differential equation similar to time independent Schr\"{o}dinger equation including a potential.
Focusing on the near of the $\mathcal{CH}$, one should be demanded
two following linearly independent solutions
\begin{align}\label{radial_solution}
\Psi^{(1)}&=e^{-i\omega u}R^{(1)}(r)Y_{\ell m}(\theta,\phi)~,
\nonumber \\
\Psi^{(2)}&=e^{-i\omega u}R^{(2)}(r)Y_{\ell m}(\theta,\phi)\left(r-r_{c}\right)^{i\omega/\kappa_{c}}~,
\end{align} for the radial second order differential equation of $R(r)$.
Here, $u$, $r_{c}$, and $\omega$  denote the retarded time coordinate, the location of the $\mathcal{CH}$, and QNM frequency, respectively. To violate SCCC, the existence of weak solutions across the $\mathcal{CH}$ is essential. 
Once  this is possible that the following integral of the non-linear field equations multiplied with a test smooth function $\psi$ over a region around the
$\mathcal{CH}$ (with volume $\mathcal{V}$), is finite \footnote{The key point is that linear perturbations may not be differentiable, at the $\mathcal{CH}$, meaning that such a perturbation can not satisfy the equations of motion at the $\mathcal{CH}$. More technical, if one treats the Klein-Gordon's equation $\square \Psi=0$ as a first-order perturbation, sourcing a second-order metric perturbation $h_{\mu\nu}^{(2)}$, finally will deal with the perturbation equation as $\mathcal{L}h_{\mu\nu}^{(2)}\equiv T_{\mu\nu}[\Psi]$. 
Here, $\mathcal{L}$ and $T_{\mu\nu}[\Psi]$ are a certain second order differential operator and the energy momentum of
the scalar field, respectively. If $\Psi$ and $h_{\mu\nu}^{(2)}$ be non-differentiable, then the relevant  perturbation equation  becomes meaningless. To get rid of this issue, it multiplies by a smooth, compactly supported test function as $\psi$, and in end integrating by parts, just similar to Eq. (\ref{In}). Indeed, the test function $\psi$ lets us to integrating by parts in the absence of boundary terms. In this way, the scalar field to address the weak solution at the $\mathcal{CH}$ should belong to the space of functions that are locally square integrable. Here the test smooth function has a vital role since a function is locally square integrable provided that it is square integrable when multiplied by any smooth test function. The essential condition on the test smooth function is that it should not zero on the $\mathcal{CH}$. To find more technical discussions, we recommend readers to see Refs. \cite{ Dias:2018ynt,Dias:2018etb}.} \cite{Mishra:2019ged}
\begin{align}\label{In}
\int_{\mathcal{V}} d^dx\sqrt{-g}\psi(\partial_{\mu}\Psi\partial ^{\mu}\Psi)~.
\end{align}
Now using the above solutions, one can determine that the integral of the kinetic term of the scalar field i.e., $(\partial_{\mu}\Psi\partial ^{\mu}\Psi)$ over the Cauchy surface, correspond to integral $(r-r_{c})^{2(i\omega/\kappa_{c}-1)}$ which with regard to the deterministic factor (\ref{beta}),
then its final form after integration reads as $(r-r_c)^{2\beta-1}$.
As a consequence, if $\beta\geq1/2$, the perturbing scalar field $\Psi(t,r,\Omega)$ is regular at the $\mathcal{CH}$, indicating that it can be extended safely beyond $\mathcal{CH}$ and subsequently violates the SCCC.

Note that the heart of this analysis is the late-time tail behavior of the scalar field outside the black hole.
In other words, the principle assumption that lets us even beyond GR consider Eq. (\ref{beta}) as a deterministic factor is that we still believe that the massless scalar perturbations at late times decay exponentially outside the region of a black hole. Except for the standard RN-dS \cite{Brady:1996za,Chambers:1997ef}
for the most other charged black holes admitted by extended Einstein-Maxwell theories, there is no rigorously proof, and people just have demanded frequently it as an accepted assumption in their works \cite{Rahman:2018oso}-\cite{Rahman:2020guv}.
However, in light of arguments released in \cite{Brady:1996za,Chambers:1997ef} and \cite{Gundlach:1993tp,Gundlach:1993tn}
one will find that the late time tail behavior of massless scalar waves (particularly for the multip oles $\ell>0$)
propagating on spherically symmetric backgrounds (in the absence or presence of cosmological constant), depends on the general form of effective potential at faraway distances. In other words, because the effective potentials for asymptotically flat and dS backgrounds at far distances respectively behavior the form of power-law and exponentially, the behavior of their corresponding massless scalar tail at the late time is such, too. Therefore, in any extended framework, by investigating the behavior of effective potential at a faraway distance, one can show whether the late time behavior of massless scalar wave on a dS background decay exponentially or not. Before the numerical analysis of PS-QNMs, we will check it separately for each of the two extended gravity models at hand to ensure the validity of Eq. (\ref{beta}).

\section{Photon Sphere-QNMs and deterministic inequality of SCCC} \label{sec.III}

Black holes are not, in essence, static objects rather they can undergo some oscillations so-called QNMs arising from the variety of perturbation fields (typically scalar, vector, and tensor) in the black hole vicinity or even spacetime itself. These oscillations indeed are some complex eigenvalues of the wave equations with quasi-normal frequencies (QNFs), which, unlike normal modes, eventually will drop exponentially due to the system's energy
loss. In this way, the black hole finally settles in a stationary equilibrium state. Reading the real and imaginary parts of QNFs, one finds the oscillation frequencies and the damping rates of the modes, respectively.
The importance of studying QNMs is that by modeling the late time behavior of perturbed black holes, they address the footprints of a back hole geometry without any dependency on how they were excited.
In other words, the QNMs are strongly dependent on the parameters involved in the back hole metric.
For instance, if we give long enough time to the black hole merging phenomenon, in the post-merger stage, we deal with a single black hole in equilibrium along with some small perturbations, indicating the existence of QNMs.
Given the close link between the QNMs of a black hole with the specific boundary conditions required at the event horizon, namely pure ingoing waves, it is prone to distinguish the black hole from its horizonless counterparts \cite{Cardoso:2016rao}.
Historically, the study of black hole QNMs as one of the hot topics in modern physics has developed from long ago
with seminal papers \cite{Regge:1957td}-\cite{Iyer:1986np} until recent years \cite{Konoplya:2003dd,Matyjasek:2017psv,Matyjasek:2019eeu} \footnote{In recent years, we have seen a great
interest in the analytically as well as numerically computation of the QNMs arising from various types of perturbation
fields within a wide class of black hole solutions, see e.g. \cite{Konoplya:2002ky}-\cite{Sun:2020sgn} and also
references therein.}. Theoretically, QNMs are effective and quantitative tools with rich usages in a deeper understanding of frameworks such as higher-dimensional black hole physics, AdS/CFT correspondence and accurate testing of the GR via probing the validity of some fundamental theorems/conjectures as no-hair and strong cosmic censorship \cite{Berti:2009kk}.

Even though the calculation of QNMs relevant for the perturbation of black holes chiefly via numerical techniques is acquired, analytical treatment is also possible under certain conditions. One of them is based on a geometrical-optics approximation which for the first time, was proposed in seminal papers \cite{Ferrari:1984ozr,PhysRevD.31.290}. The main idea of this method is the null geodesics trapped at the unstable photon orbit. More technically, in this method, the real and imaginary parts of the QNMs, are given respectively by the angular frequency of rotation of a photon in the \emph{``photon circular orbit''} (PCO) and the largest Lyapunov exponent measuring the growth of the perturbation surrounding PCO, see \cite{Ferrari:1984ozr,PhysRevD.31.290} and \cite{Cornish:2003ig,Cardoso:2008bp,Konoplya:2017wot}.
The salient feature of this method is that its outputs are in a nice agreement with numerical methods in the eikonal limit ($\ell\gg1$) \cite{Hod:2009td,Konoplya:2017wot}. To see the connection between the Lyapunov exponent and the growth of perturbation around PCO, we write down the equation for radial null geodesics in the equatorial plane ($\theta=\pi/2$) as
\begin{equation}\label{rdotVr}
\dot{r}^{2}=V_{\rm eff}(r)~,
\end{equation}
 where symbol `dot' and $V_{\rm eff}$ respectively denote the
derivative with respect to the affine parameter relevant to the null geodesics and
effective potential associated with the radial null geodesics.
Also, the PCO with radius $r_{\rm ph}$, is obtained by solving the following equation
\begin{equation}\label{rph}
V_{\rm eff}(r_{\rm ph})=0=V_{\rm eff}'(r_{\rm ph})~.
\end{equation} 
Since in the position of $r_{\rm ph},$  the value of the effective potential is maximum, then PCO is unstable.
Hence, the Lyapunov exponent $\lambda$ is associated with infinitesimal fluctuations near to the PCO, i.e., $r=r_{\rm ph}+\delta r$, with the assumption that $\delta r$ is a small perturbation. Inserting this expression and then expanding the right hand side of (\ref{rdotVr}) around the PCO, we come to an expression as
\begin{equation}\label{prelya}
\begin{aligned}
\left(\dot{\delta r}\right)^{2}=\frac{1}{2}V_{\rm eff}''(r_{\rm ph})\delta r^{2}~,
\end{aligned}
\end{equation} 
for the time evolution of the perturbed quantity $\delta r$. Note that here, we have kept terms up to quadratic order in the expansion of the right hand side of (\ref{rdotVr}). To get rid of the affine parameter, we divide both sides of (\ref{prelya}) with $\dot{t}^{2}$, which results in the following solutions
\begin{equation}\label{lya}
\delta r=A \exp(\pm \lambda t)~,
\end{equation} 
with constant of integration $A$ and the Lyapunov exponent $\lambda$ yielding the growing (+ sign) and decaying (- sign) rates of the PCO with the following analytical expression \cite{Cardoso:2008bp}
\begin{equation}\label{lyapunov}
\lambda=\sqrt{\dfrac{V_{\rm eff}''}{2\dot{t}^{2}}}\bigg|_{r=r_{\rm ph}}~.
\end{equation} Finally, it is shown that in the eikonal limit, the Lyapunov exponent connects to the imaginary part of the QNMs as
\begin{equation}\label{imomega}
\operatorname{Im}(\omega)=-\left(n+\frac{1}{2}\right)\lambda~,~~~n=0,1,2,...
\end{equation} where $n$ is the overtone number. Since the frequencies of the QNMs appear as $\exp(-i\omega t)$, it will decay at a faster rate as the value of $n$ gets bigger. It means that the longest lived mode corresponds to $n=0$ i.e.,
$\{\textrm{Im}(\omega)\}_{\rm min}=-\lambda/2$. As a result, in the case of admitting a  $\mathcal{CH}$ by spacetime, one can certainly calculate the relevant surface gravity $\kappa_{c}$, and subsequently, the parameter $\beta$ ( Eq. (\ref{beta})), in the eikonal limit takes the following form
 \begin{equation}\label{sccviolation}
\beta_{\rm PS}=\frac{\lambda}{2\kappa_{c}}~.
\end{equation} 
 As stressed before, the parameter $\beta$ is a deterministic factor clarifying the validity domain of the SCCC in diverse black hole spacetimes.

In the following, we will calculate the Lyapunov exponent and subsequently $\beta$ in the eikonal limit for both static and spherically symmetric black hole spacetime. In this regard, we take the following static and spherically
symmetric metric
\begin{equation}\label{staticspherically}
ds^{2}=-N(r)dt^{2}+N(r)^{-1}dr^{2}+r^{2}d\Omega^{2}_{2}~,
\end{equation} with the line element of the $2$-sphere $d\Omega^{2}_{2}=\sin\theta^2 d\theta^2+d\varphi^2$ and arbitrary lapse function $N(r)$ which, in essence, comes from solving the associated gravitational field equations related to any gravity theory. Therefore, the results extracted from the above metric are applicable in any context which gives rise to a static and the spherically symmetric solution. By restricting the motion of a particle on the equatorial plane, then the associated Lagrangian reads as
 \begin{equation}\label{lagrangian}
\mathcal{L}=\frac{1}{2}\Big(-N(r)\dot{t}^{2}+N(r)^{-1}\dot{r}^{2}+r^{2}\dot{\phi}^{2}\Big)~,
\end{equation} where `dot' refers to the  derivative with respect to proper time, proper length, and the affine parameter in the context of timelike, spacelike, and null trajectories, respectively. Given that the metric at hand has no explicit dependence on the coordinates $t$ and $\phi$, thereby, the corresponding conjugate momentums of trajectory i.e., the energy $p_{t}=-E$ and angular momentum $p_{\phi}=L$, are constants of motion. As a result, the geodesic equation for the radial coordinate just is non-trivial and written as
\cite{Cardoso:2008bp}
\begin{align}\label{geodesicnospin}
\dot{r}^{2}=E^{2}-N(r)\left(\frac{L^{2}}{r^{2}}-\epsilon \right)~,
\end{align} where $\epsilon=g_{\mu\nu}u^{\mu}u^{\nu}=(-1,0,1)$ respectively for timelike, null and spacelike geodesics. Due to this fact that the determination of Lyapunov exponent depends explicitly on the PCO,
so by setting $\epsilon=0$ into (\ref{geodesicnospin}), the radial null geodesics satisfy the following equation
\begin{equation}\label{potentialnospin}
\dot{r}^{2}=E^{2}-N(r)\frac{L^{2}}{r^{2}}\equiv V_{\rm eff}(r)~.
\end{equation} 
Putting the above potential into conditions released in (\ref{rph}), one can
immediately come to the following equations
\begin{equation}\label{rphnospin}
\begin{aligned}
\frac{E^{2}}{L^{2}}=\frac{N(r)}{r^{2}}~,~~~~~~
\frac{N'(r)}{N(r)}=\frac{2}{r}~,
\end{aligned}
\end{equation} 
where ``prime" is indeed derivative with respect to $r$. Now,
by specifying the potential from (\ref{potentialnospin}), one able to
compute $V_{\rm eff}''(r)$ necessary to derive the Lyapunov exponent.
Finally, by combining (\ref{lyapunov}) and (\ref{rphnospin}) plus the relevant
expression for $V_{\rm eff}''(r)$ on the PCO ($r=r_{\rm ph}$), one can arrive at the Lyapunov exponent
\begin{equation}\label{lyanospin}
\lambda =\sqrt{\left(\frac{N(r_{\rm ph})}{r_{\rm ph}}\right)^2-\frac{N(r_{\rm ph})N''(r_{\rm ph})}{2}}~,
\end{equation} for the static and spherically symmetric spacetime
(\ref{staticspherically}). Note that in (\ref{lyapunov}) due to this fact that $p_{t}=-E$ then $\dot{t}$ replaced
with $\frac{E}{N(r)}$.
Now by having the above expression for the Lyapunov exponent, one can look for the possibility of violation of SCCC in a static and spherically symmetric spacetime (\ref{staticspherically}) with any arbitrary choice of the lapse function $N(r)$, including two Cauchy and event horizons.
They are indeed corresponding to the smallest and biggest roots
$(r_{c},r_{e})$ of the equation $N(r)=0$, respectively. Eventually, by regarding the surface gravity relevant to the
$\mathcal{CH}$ i.e.,  $\kappa_{c}=\frac{N'(r_{c})}{2}$, then the parameter $\beta_{\rm PS}$ for QNMs in the eikonal limit
(i.e., large $\ell$), reads as
\begin{equation}\label{betastatic}
\beta _{\rm PS}=\sqrt{\left(\frac{N(r_{\rm ph})}{r_{\rm ph} N'(r_{\rm c})}\right)^2-\frac{N(r_{\rm ph})N''(r_{\rm ph})}{2N'(r_{c})^{2}}}~.
\end{equation}
The above expression is valid for any static and spherically symmetric spacetime plus a positive cosmological constant which indicates a non-trivial $\mathcal{CH}$, in addition to the event horizon.

In the rest of the paper, by focusing on PS-QNMs, we will conduct numerically the formalism proposed above for the charged-dS black hole solutions, which arose from two extended gravity theories: $f(R)$ and EMSG. In this way, modified gravity gives us a richer framework to finding $\beta _{\rm PS}<1/2$, include a free model parameter in addition to the electric charge and cosmological constant.
Based on the modern studies such as \cite{Hintz:2016gwb,Dafermos:2012np,Cardoso:2017soq,Dias:2019ery,Hod:2018dpx,Cardoso:2018nvb} for violation of SCCC, definitely must $\beta$-criterion for all of QNMs attributed to black hole solution becomes larger than half. This means that in case of introducing a counter-example, via finding regions with $\beta _{\rm PS}<1/2$, even for one of the QNMs, we can reject the claim of violation of the SCCC. 
In this way, the validation of SCCC is restored. 
This is something we will show explicitly in what follows.

\section{SCCC for charged black holes in $f(R)$-gravity} \label{sec.FR}

By utilizing the formalism discussed  above we here will investigate SCCC within the scope of a charged static and spherically
symmetric spacetime admitted by the Einstein-Maxwell-$f(R)$ gravity
\begin{eqnarray}\label{sf}
S =\frac{1}{2} \int \text{d}^{4}x\sqrt{\mid g\mid}\,\bigg(R+f(R)-F_{\mu\nu}F^{\mu\nu}\bigg)\,,
\end{eqnarray} with $F_{\mu\nu}=\partial_\mu\,A_\nu-\partial_{\nu}A_\mu$ in which $A_\mu$ is the potential four-vector.
By taking the static spherically symmetric metric (\ref{staticspherically}), the
black hole solution with constant curvature $R_0$, takes the following form  \cite{delaCruzDombriz:2009et,Moon:2011hq,delaCruzDombriz:2012xy}
\begin{equation}\label{fr0}
N(r)=1-\frac{2M}{r}+\frac{1}{1+f'(R_0)}\frac{Q^2}{r^2}-\frac{\Lambda}{3}r^2\,,
\end{equation} where $f'(R_0)\equiv(\frac{d f(R)}{d R})_{R_0}$.
The above solution is different from its counterpart in GR by a multiplicative factor $1+f'(R_0)$, which needs to be positive  to maintain gravitational and thermodynamical viability conditions \cite{Cembranos:2011sr}. Thus,  the values of $f'(R_0)$ are limited within ranges:  $-1<f'(R_0) <0$ and $f'(R_0)>0$.
For both cases, the above metric discloses the presence of a timelike singularity at the center of the black hole, just like its standard counterpart (RN).
Note that the factor $1+f'(R_0)$ is induced by spacetime, and it is expected that in high curvature regions, shows phenomenological imprints separable from the electric charge $Q$. More exactly, one is not allowed to label $\frac{Q}{\sqrt{1+f'(R_0)}}$ as an electric charge because the four-vector potential $A_\mu$ takes the form $\frac{Q}{r}(-1,0,0,0)$ \cite{delaCruzDombriz:2010xy,delaCruzDombriz:2012xy,Khodadi:2020cht}, meaning that $Q$ is responsible for the electrical charge.
Hence, the factor $1+f'(R_0)$ is an additional effect arising from spacetime which is expected to be significant just at close distances to the black hole. Given that here $Q$ plays the role of electric charge, not $\frac{Q}{\sqrt{1+f'(R_0)}}$,  depending on value of the $f'(R_0)$, the value of the charge at which the two horizons merge is larger and smaller than the standard value in GR. More clearly, the first case occurs if $-1<f'(R_0) <0$, while for the second case it is require that $f'(R_0)>0$. This result explicitly suggests the possibility of screening and anti-screening effects of the electric charge for two regions negative and positive of the model parameter $f'(R_0)$\footnote{It would be interesting to know that solar system tests severely have constrained the absolute value of $\mid f'(R_0)\mid$ from above, $\lesssim10^{-6}$, see Ref. \cite{Hu:2007nk}. Combining the solar system test and the violation of the equivalence principle has resulted in more severe constraints \cite{Capozziello:2007eu}.}. 

\begin{figure}
	\minipage{0.6\textwidth}
	\includegraphics[width=\linewidth]{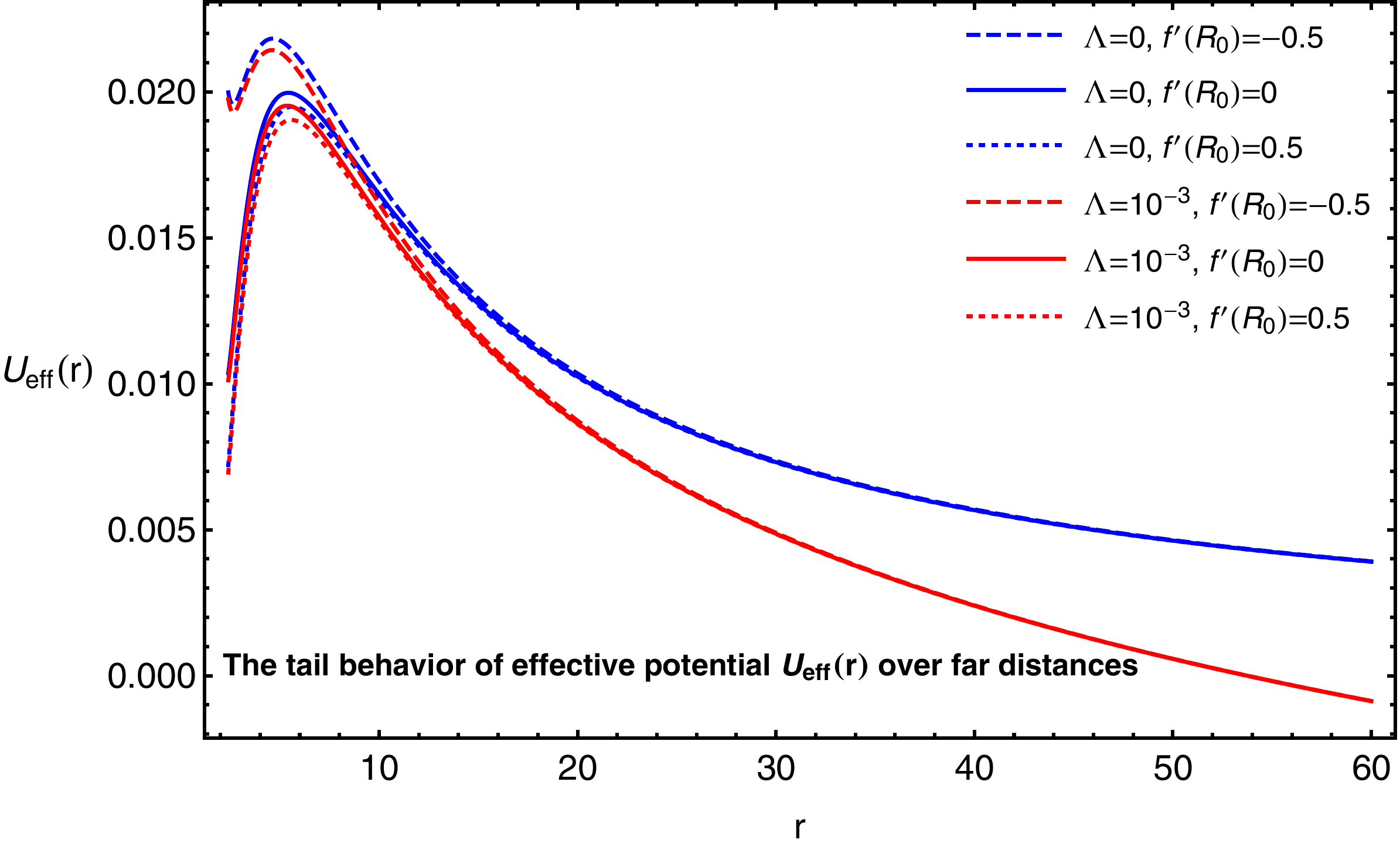}
	\endminipage\hfill
	\caption{ The tail behavior of effective potential  (\ref{potf}) for the $f(R)$-model over faraway distances in the absence (blue-solid, dashed, dotted) and presence (red-solid, dashed, dotted) of cosmological constant $\Lambda$. We set numerical values $\ell=10,~Q=0.7$ and $M=1$. The general behavior is not sensitive to values $Q$ and $f'(R_0)$.}
	\label{potfig1}
\end{figure}

Before moving on to QNMs analysis, let us check the accuracy of $\beta$-criterion (\ref{beta}) via confirming
this point that the rate of decay of the tail of the massless scalar waves at late times on the top of dS-spacetime
is faster relative to the flat asymptotically counterpart.
By following the discussions in \cite{Brady:1996za,Chambers:1997ef,Gundlach:1993tp,Gundlach:1993tn} (as already mentioned),
this can do by means of tracking the long distance behavior of relevant effective potential.
In other words, we expect the tail of effective potential in the presence of a tiny positive cosmological constant decay in faraway distances faster than when  $\Lambda=0$.
Concerning the scalar perturbations over the black hole background which are governed by the
Klein-Gordon's equation $\frac{1}{\sqrt{-g}}\partial_{\mu}(\sqrt{-g}g^{\mu\nu}\partial_{\nu}\Psi)=0$, then the general form of effective potential reads as \footnote{Generally in the second term into parenthesis, there is a coefficient as $(1-s^2)$ which comes from the spin of the relevant perturbation field.
However, concerning PS-modes due to the focus on the eikonal limit,
 the effective potential is dominated by the $\ell(\ell+1)$, meaning that the role of spin is negligible. As a result, for the PS-QNMs one should expect to have similar QNFs, independent of the type of the perturbation fields.}
\begin{equation}\label{potf}
U_{eff}(r)=N(r)\bigg(\frac{\ell(\ell+1)}{r^2}+\frac{N'(r)}{r}\bigg)\,.
\end{equation}
By taking the metric solution (\ref{fr0}) into (\ref{potf}) and expanding around infinity, then, the effective potential
in faraway distances behaves as shown in Fig. (\ref{potfig1}).
There are two clear points in this figure.
First,  as the distance increases, the long distance behavior of $U_{eff}(r)$ in the context of Einstein-Maxwell-$f(R)$ gravity, is identical to its standard counterpart. In this way, one can easily check that value of $f'(R_0)$ has virtually no role in infinity distances.
Second, in the presence of a cosmological constant, the behavior of the tail of effective potential in faraway distances is no longer in the form of power-law, rather decays faster like exponentially. The presence of the cosmological constant increases the slope falling the tail of potential in far distances, just as expected.

\begin{figure}
	\minipage{0.3\textwidth}
	\includegraphics[width=\linewidth]{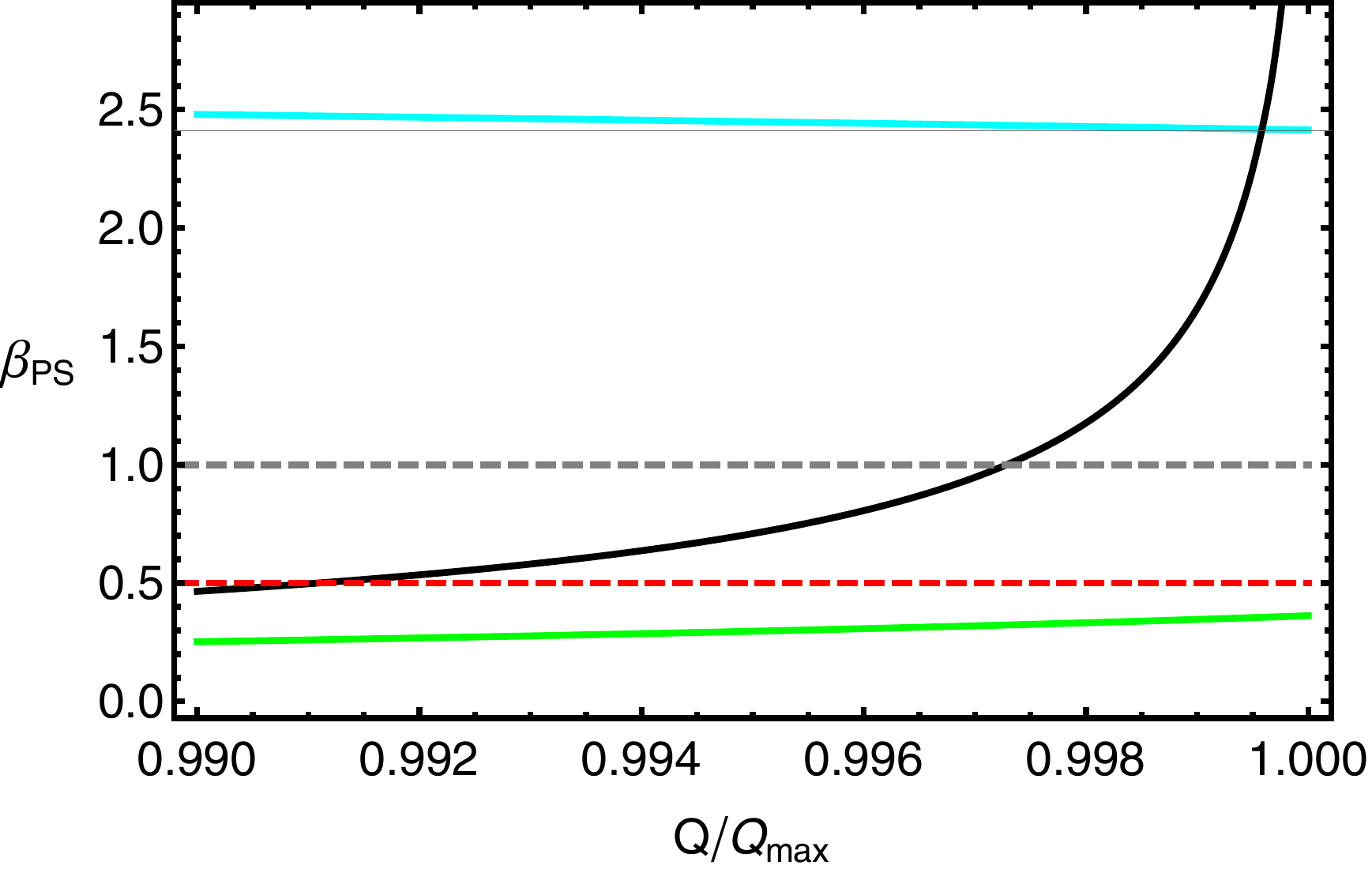}
	\endminipage\hfill
	\minipage{0.3\textwidth}
	\includegraphics[width=\linewidth]{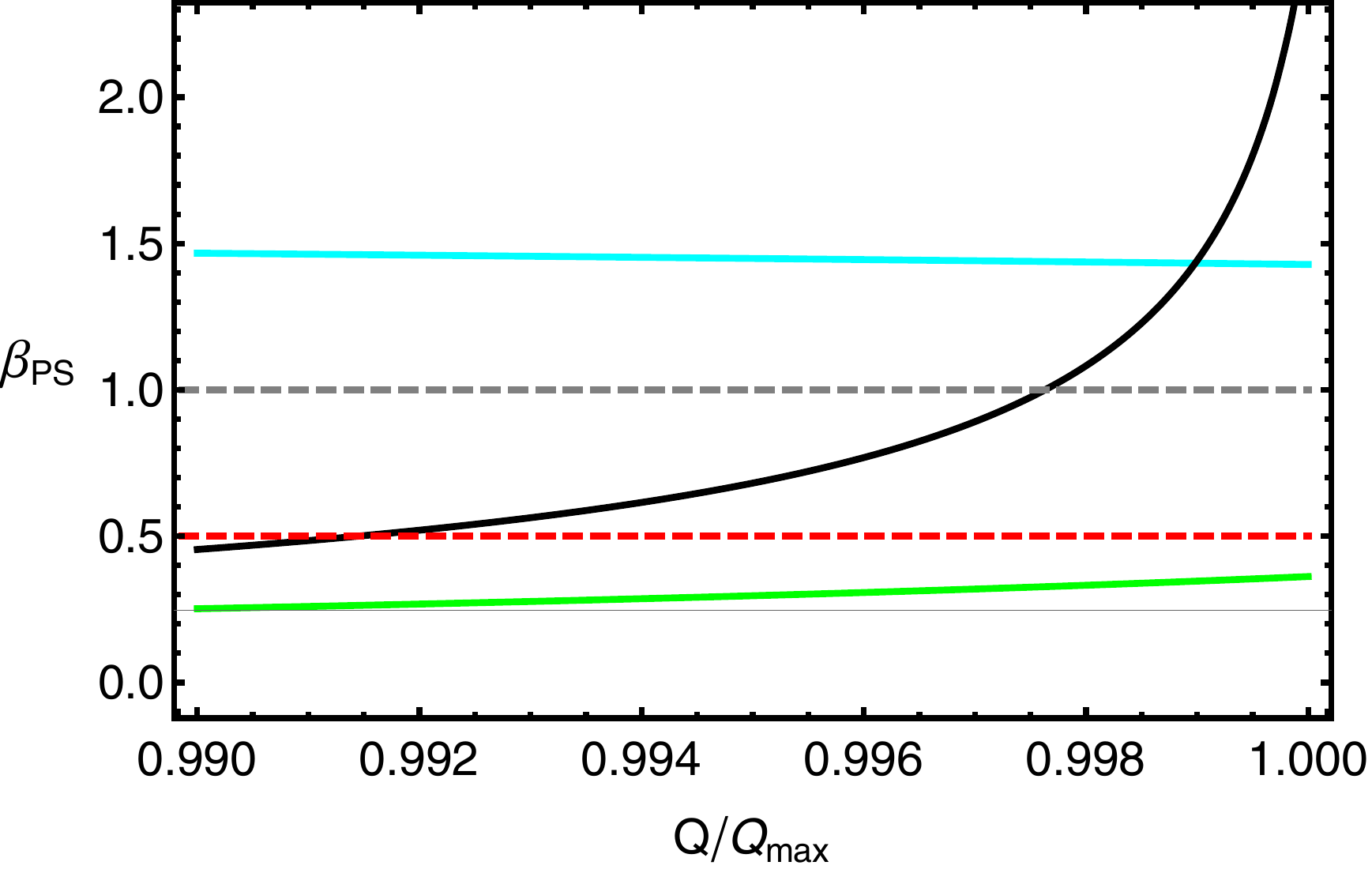}
	\endminipage\hfill
	\minipage{0.3\textwidth}%
	\includegraphics[width=\linewidth]{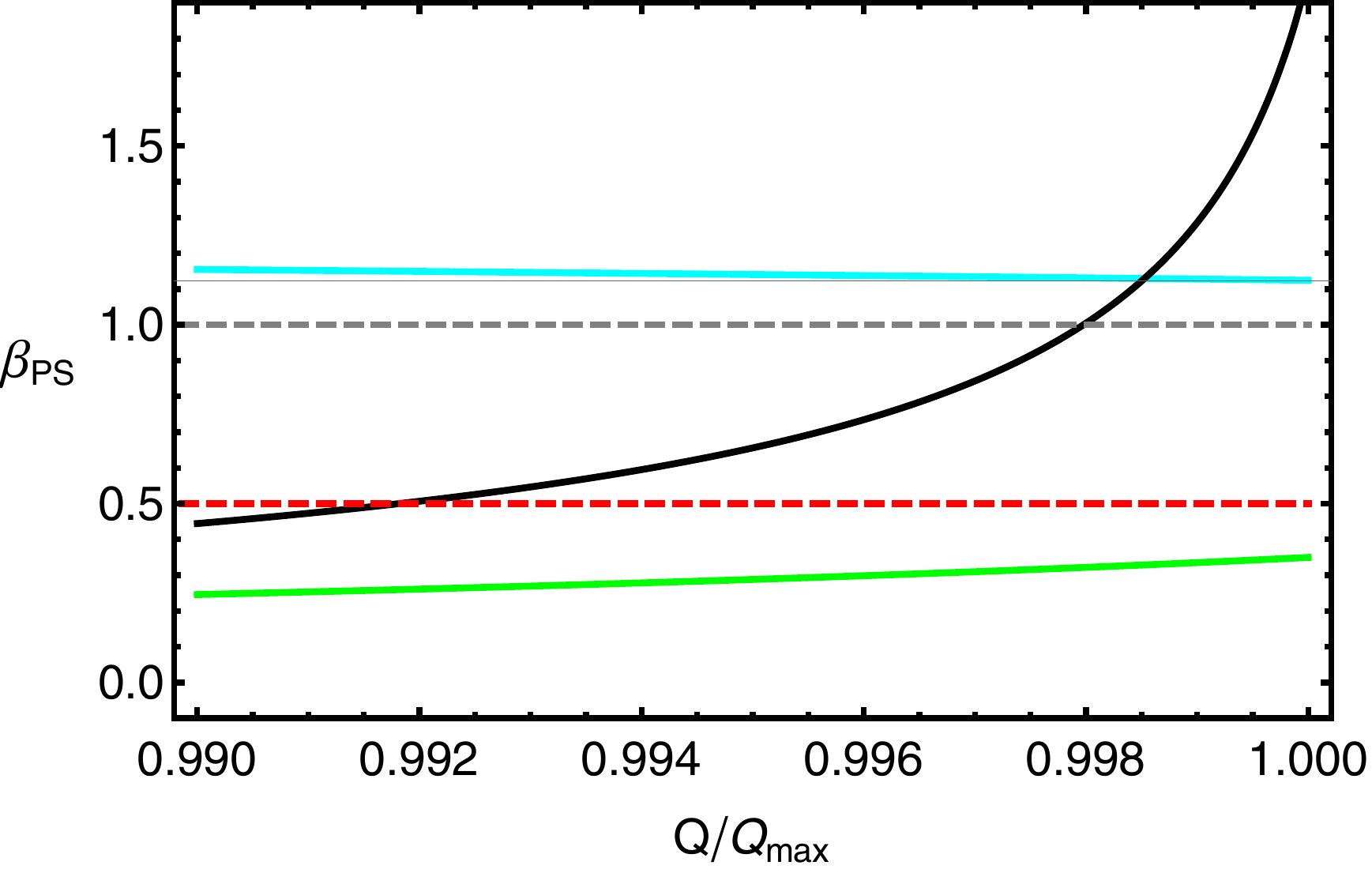}
	\endminipage\hfill
	\minipage{0.3\textwidth}
	\includegraphics[width=\linewidth]{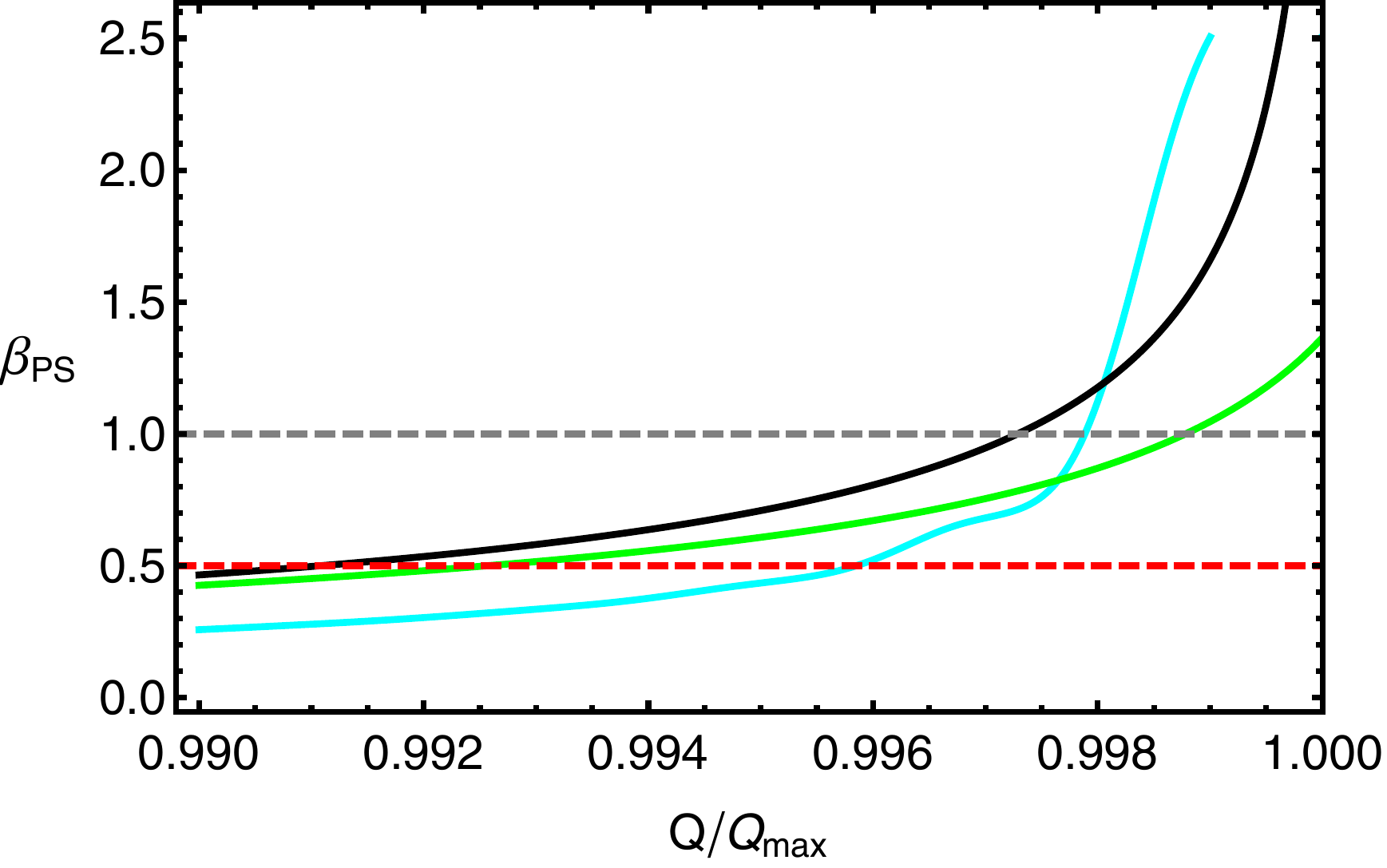}
	\endminipage\hfill
	\minipage{0.3\textwidth}
	\includegraphics[width=\linewidth]{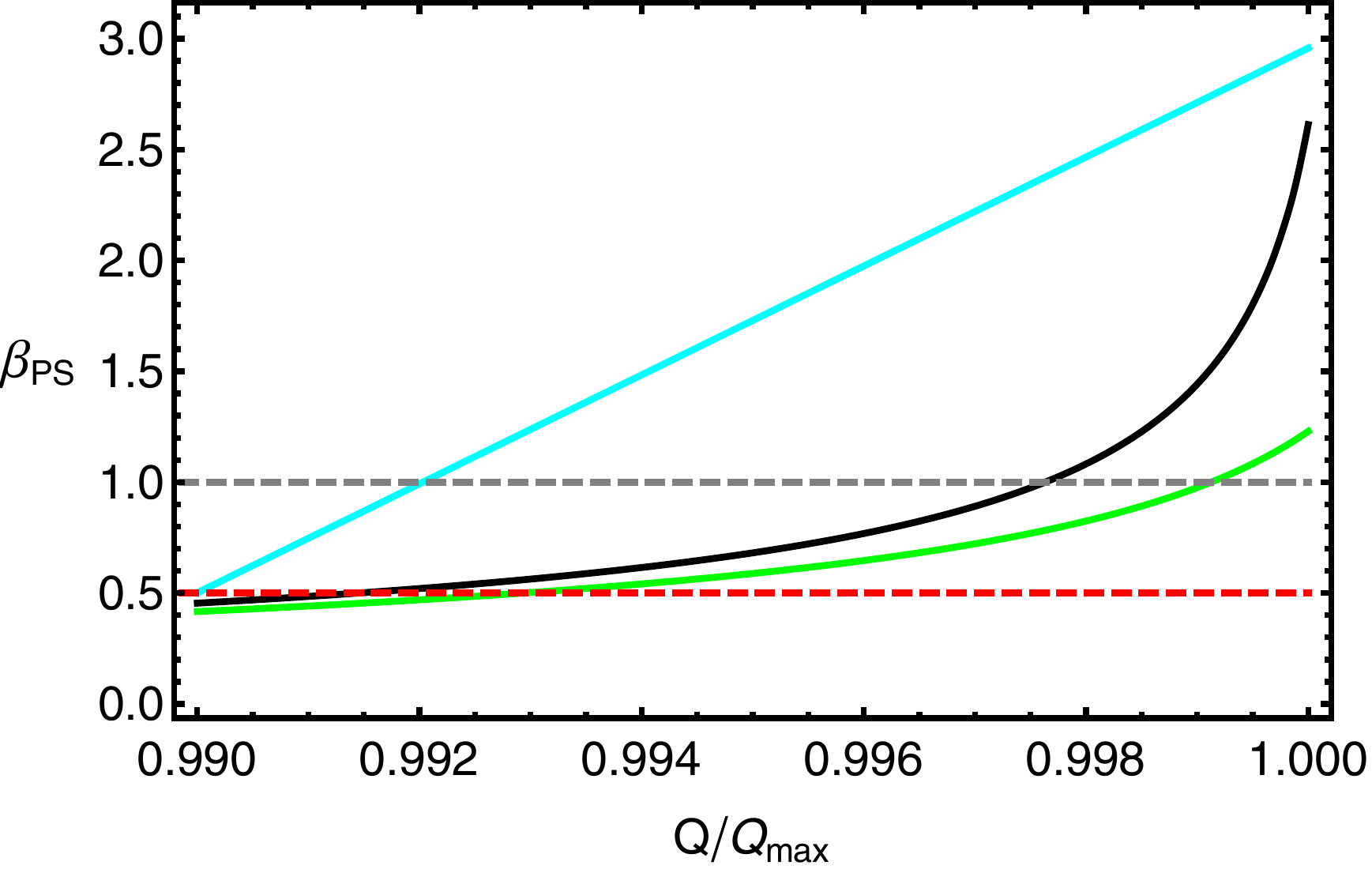}
	\endminipage\hfill
	\minipage{0.3\textwidth}%
	\includegraphics[width=\linewidth]{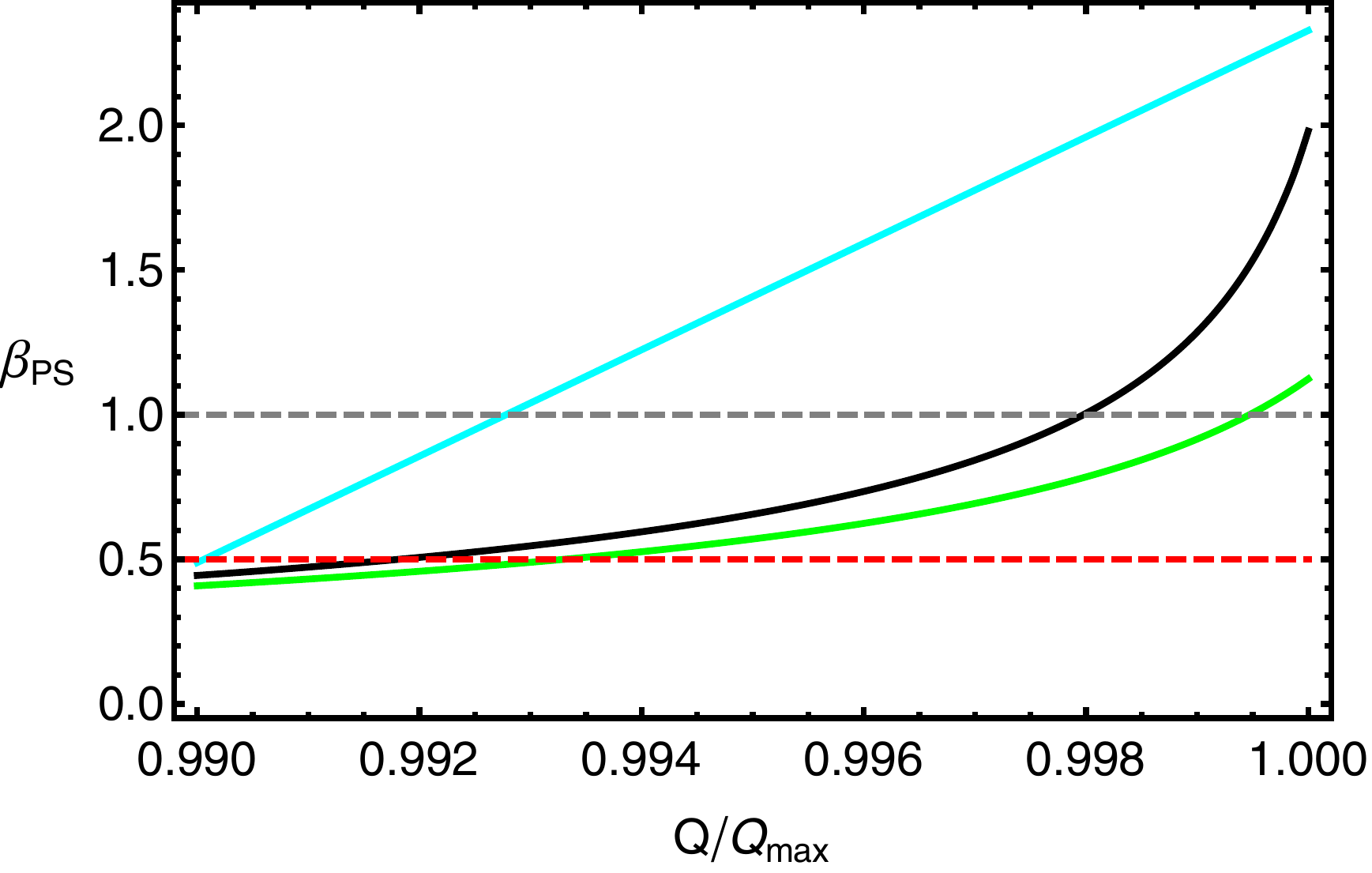}
	\endminipage\hfill
	
	\caption{ The variation of $\beta$ arising from PS-QNMs in terms of electrical charge $Q$ for different values of 	$f'(R_0):-3\times10^{-2},0,3\times10^{-2}$ (first \textcolor{blue}{ row}) and $-3\times10^{-3},0,3\times10^{-3}$ (second \textcolor{blue}{ row}) respectively correspond to curves: cyan, black and green with fixed values $\Lambda: 10^{-3}, 3\times10^{-3}, 5\times10^{-3}$ in any \textcolor{blue}{ row} from left to right. Note that throughout this manuscript the $\beta$-plots have two common features: first, the Mass is normalized to unity ($M=1$), second, the horizontal red and gray dashed lines show $\beta=1/2$ (boundary line of SCCC violation), and $\beta=1$, respectively.}
	\label{Modeff}
\end{figure}

\begin{figure}
	\minipage{0.5\textwidth}
	\includegraphics[width=\linewidth]{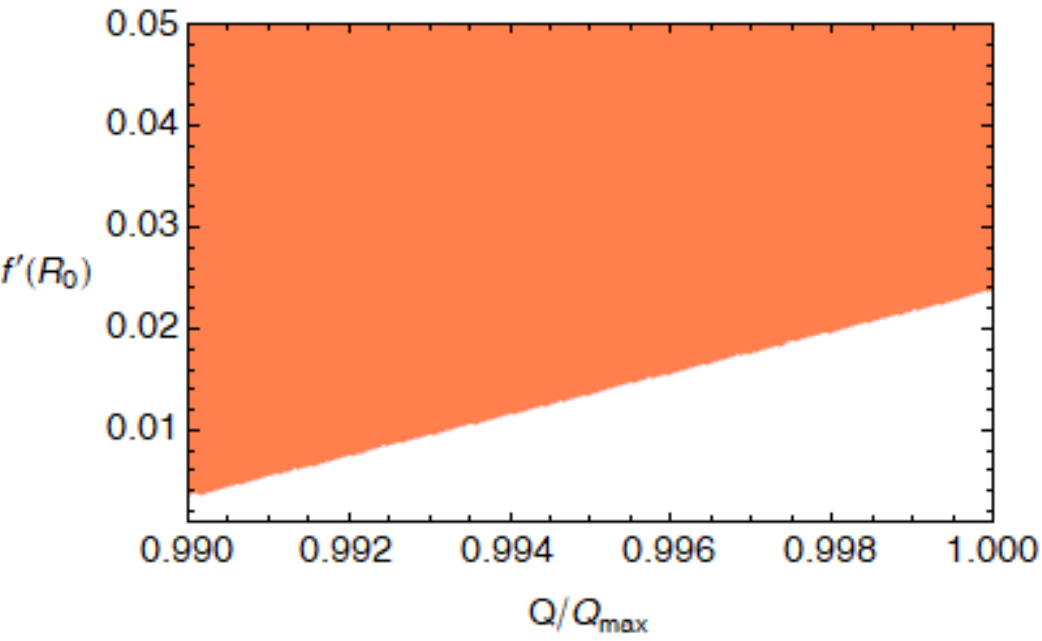}
	\endminipage\hfill
	\minipage{0.5\textwidth}
	\includegraphics[width=\linewidth]{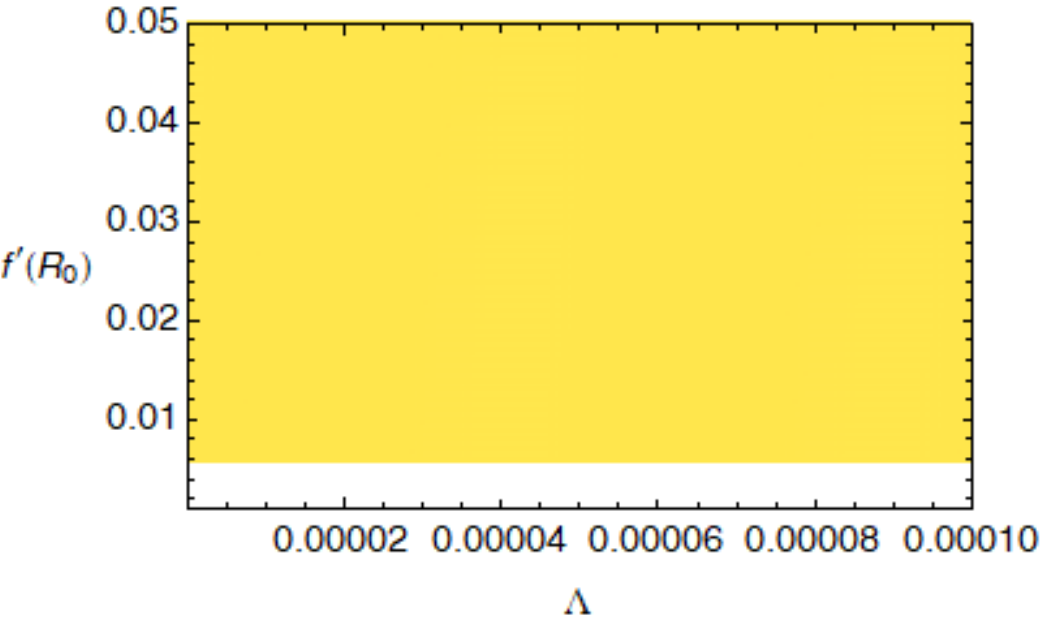}
	\endminipage\hfill
	\caption{ The colored-shaded area in the left and right panels are respectively the allowed region
of $Q/Q_{max}-f'(R_0)$  and $\Lambda-f'(R_0)$ parameter spaces in which the $\beta$-criterion
related to PS-QNMs is placed within $0<\beta_{PS}<1/2$.
 Note that in order to have more resolution
on the lower limit of $f'(R_0)$, we restricted its upper limit to a given value while it can be extended to bigger values, too. The colorless region is actually the exclusion region.}
\label{Scan1}
\end{figure}
Now we start our analysis for the PS-QNMs. Concerning the Lyapunov exponent $\lambda _{\rm f(R)}$
and $\beta_{\rm PS}$, one needs to find the location of PCO ($r_{\rm ph}$) and relevant horizons ($r_{c,e}$).
Given Eq. (\ref{rphnospin}) and $N(r)=0$ one can respectively obtain them
by solving algebraic equations
\begin{equation}\label{rpheff}
\begin{aligned}
&r^2-3Mr+\frac{2 Q^2}{1+f'(R_0)}=0~, \\
&\frac{\Lambda}{3}r^4-r^2+2Mr-\frac{Q^2}{1+f'(R_0)}=0~.
\end{aligned}
\end{equation}
By putting the metric solution (\ref{fr0}) into (\ref{lyanospin}) and
(\ref{betastatic}), the Lyapunov exponent $\lambda _{\rm f(R)}$ and deterministic factor $\beta^{\rm f(R)}_{\rm PS}$, reads as
\begin{equation}\label{lyabrane}
\lambda _{\rm f(R)}=\sqrt{-\frac{\left((1+f'(R_0)) r_{\text{ph}}^2-2 Q^2\right) \left((1+f'(R_0)) \left(6
		Mr_{\text{ph}}+\Lambda  r_{\text{ph}}^4-3 r_{\text{ph}}^2\right)-3 Q^2\right)}{3(1+f'(R_0))^2
		r_{\text{ph}}^6}}~.
\end{equation}
and
\begingroup\makeatletter\def\f@size{10}\check@mathfonts
\begin{equation}\label{betaeff}
\begin{aligned}
&\beta^{\rm f(R)}_{\rm PS}=\frac{\sqrt{-\frac{ M}{r_{\text{ph}}}+\frac{Q^2}{2(1+f'(R_0)) r_{\text{ph}}^2}-\frac{\Lambda  r_{\text{ph}}^2}{6}+\frac{1}{2}}}{-\frac{2 M}{r_c^2}+\frac{2 Q^2}{(1+f'(R_0)) r_c^3}+\frac{2 \Lambda  r_c}{3}} \times  \\ \nonumber
&\sqrt{-\frac{4 \bigg((1+f'(R_0))
		\left(\Lambda  r_c^4-3 M r_c\right)+3 Q^2\bigg) \left((1+f'(R_0)) \left(6 Mr_{\text{ph}}+\Lambda  r_{\text{ph}}^4\right)-9 Q^2\right)}{9 (1+f'(R_0))^2 r_c^3
		r_{\text{ph}}^4}+\frac{2}{r_{\text{ph}}^2}-\frac{4 M}{r_{\text{ph}}^3}+\frac{2 Q^2}{(1+f'(R_0)) r_{\text{ph}}^4}-
	\frac{2 \Lambda }{3}}~,
\end{aligned}
\end{equation}
\endgroup
respectively. These two quantities in the context of $f(R)$ modified gravity depend explicitly on the model parameter $f'(R_0)$, in addition to the Maxwell charge $Q$ and cosmological constant $\Lambda$.
Inspired by the fact that to keep SCCC, it is sufficient that one type of QNMs does not satisfy the condition $\beta\geq1/2$, so we are interested in seeing whether, in a parameter space consisting of these, we can find $0<\beta_{PS}<1/2$.
To do that, in Fig. (\ref{Modeff}) plotted the dimensionless deterministic factor $\beta_{PS}$ related to PS-QNMs against the dimensionless parameter $Q/Q_{max}$ ($Q_{max}$ is the maximum electric charge), for different choices of $f'(R_0)$ and the cosmological constant $\Lambda$.

Plots in this figure, openly expose the result of the mentioned extra effect on $\beta_{PS}$ for several fixed cosmological constants.
As evident from the two rows in Fig. (\ref{Modeff}) by taking some positive fixed values for $f'(R_0)$, the region of interest ($0<\beta_{PS}<1/2$)
is found for near-extremal electric charges and small values of the cosmological constant. While this favor region of $\beta$ is not achievable for no combination of $f'(R_0)\leq0$ and $\Lambda$.
By doing the scan of the underlying parameters, as shown in  Fig. (\ref{Scan1}), we extract some allowed regions which let the deterministic factor related to PS-modes put into $0<\beta_{PS}<1/2$.
Namely, by restricting the value of free parameter $f'(R_0)$ within some given intervals, it is then possible to control the rate of decay
of the scalar field perturbation outside the black hole to maintain the SCCC. Consistently, one can see that $f'(R_0)=0$ (RN-dS solution) belong to exclusion region. The right panel in  Fig. (\ref{Scan1}) nicely reveals that by taking into account $f'(R_0)>0$, one should no longer worry about the value of the cosmological constant. Namely, to have unstable $\mathcal{CH}$ within a given range of $f'(R_0)>0$, one can set any optional value for $\Lambda$.
This window opened to rescue SCCC is indeed a direct result of the enrichment of the charged-dS black hole solution by Einstein-Maxwell-$f(R)$ theory.

\section{SCCC for charged black holes in Energy-Momentum Squared Gravity}
 \label{sec.EMSG}
\begin{figure}
	\minipage{0.5\textwidth}
	\includegraphics[width=\linewidth]{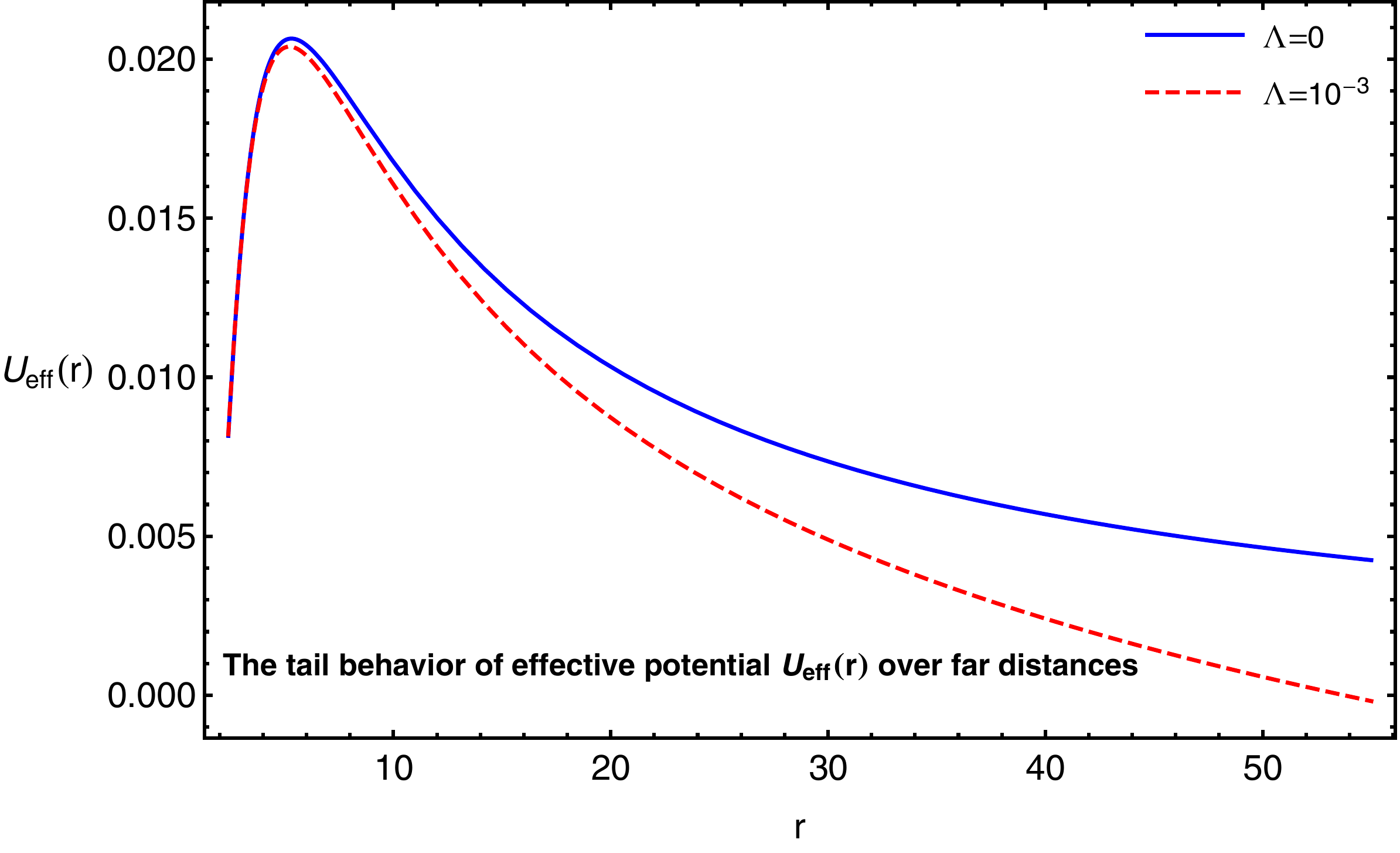}
	\endminipage\hfill
	\caption{ The tail behavior of effective potential (\ref{potf}) for the EMSG model over faraway distances in the absence (blue) and presence (red-dashed) of cosmological constant $\Lambda$. We set numerical values $\ell=10,~Q=0.7$ and $M=1$. The general behavior is not sensitive to values $Q$ and $\eta$.}
	\label{potfig2}
\end{figure}

As the second modified gravity model under our attention for the survey of SCCC within it,
we pick up the Energy-Momentum Squared Gravity (EMSG) extension of Einstein-Maxwell theory
\cite{Roshan:2016mbt}
\begin{equation}\label{action}
	S_{EMSG}=\frac{1}{2}\int \sqrt{-g}\big(R-2\Lambda -\eta (T_{\mu \nu }T^{\mu \nu })
	-\frac{1}{2}F_{\mu\nu}F^{\mu\nu}\big)\ d^{4}x.
\end{equation}
Unlike the previous model, here the generalization comes from the higher-order term of the form $T_{\mu \nu }T^{\mu \nu }$ to the matter Lagrangian. The charged static and spherically black hole solution in this model reads as
\begin{equation}\label{SG0}
	N(r)\simeq 1-\frac{2 M}{r}+\frac{Q^2}{r^2}- \frac{2\eta}{5}.\frac{Q^4}{r^6} -
	\frac{\Lambda}{3}r^2~.
\end{equation}
As one can be seen the behavior of EMSG in high dense mediums like a
black hole deviates from GR by a free parameter $\eta$. Depending on the sign of this parameter, we deal with two different causal structures for the metric inside the black hole. For $\eta >0$ this metric has a spacelike singularity at the center
of the black hole while for $\eta < 0$ it turns into a timelike singularity.
Although the EMSG with $\eta>0$ is prone to exposing a well-behaved cosmology in the late time \cite{Roshan:2016mbt} as well as the early universe \cite{Khodadi:2022zyz} \footnote{ It would be interesting to note that in the framework of inflationary cosmology, the numerical analysis done on the perturbation parameters of the EMSG model has shown that it can be a viable potential candidate from the view of Planck2018 data \cite{Faraji:2021laz}. Note that this does not mean that the modified gravity model at hand is problem-free. For instance, one can be mentioned the standard early-time singularity issue which in this  model there is no consensus on fixing it, see Refs. \cite{Roshan:2016mbt,Barbar:2019rfn,Khodadi:2022zyz}. A more important problematic one in EMSG as a subclass of modified gravities with non-minimal coupling between matter and geometry is that the conservation law of the original energy-momentum tensor is no longer guaranteed, e.g., see \cite{Sharif:2017sey,Akarsu:2017ohj}. More clearly, the perfect fluid, which in cosmology can describe effectively the behavior of Hubble flow ranging from inflation to dark energy epochs, in the framework of EMSG no longer fulfills the standard covariant energy-momentum conservation. It is also worth mentioning that in EMSG-based cosmology, the cosmological constant issue at the late time remains unsolved, too \cite{Akarsu:2018drb}. }, 
however, the possibility of $\eta<0$ does not rule out by constraints obtained from astrophysics objects with a high density such as neutron stars \cite{Akarsu:2018zxl}. This was also recently confirmed by studying the motion of light in the weak-field limit of EMSG \cite{Nazari:2022fbn}. Therefore,  by focusing on PS-QMs we will survey the possibility of revival of the SCCC for both cases $\eta>0$ and $\eta<0$. Before that, we have to check the accuracy of $\beta$-criterion (\ref{beta}) via the same argument proposed already in model $f(R)$.
In this regard, by putting the metric solution (\ref{SG0}) into (\ref{potf}) and expanding around infinity, then, the effective potential in far distances behaves as shown in Fig. (\ref{potfig2}). Since the correction arising from EMSG is of the form $1/r^6$, over long distances, deviation from standard GR is not detachable, as one can see in Fig. (\ref{potfig2}). So here also, like RN-dS, the tail of the scalar field at late times drops at a faster rate relative to $\Lambda=0$.

\begin{figure}
	\minipage{0.32\textwidth}
	\includegraphics[width=\linewidth]{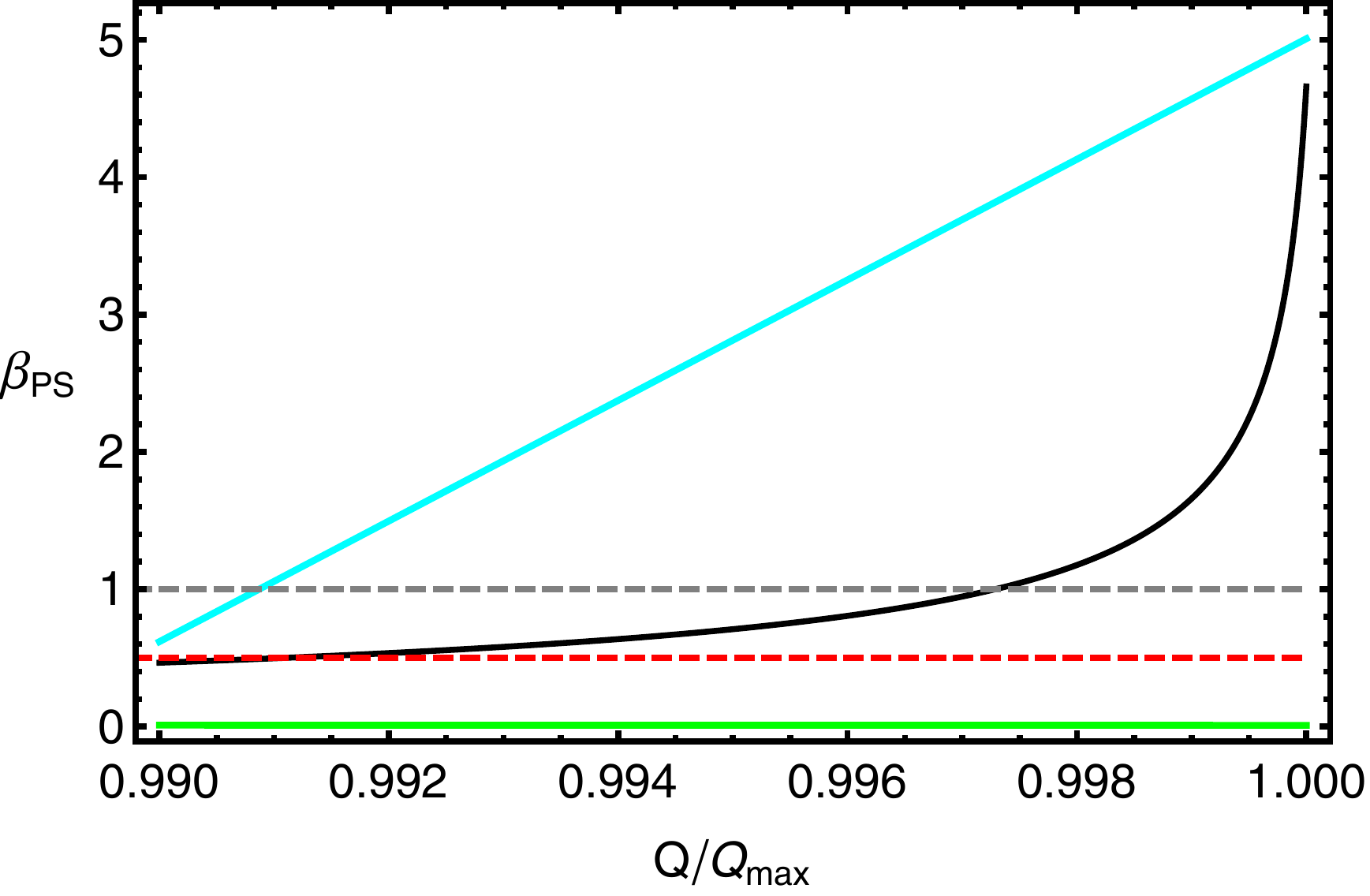}
	\endminipage\hfill
	\minipage{0.32\textwidth}
	\includegraphics[width=\linewidth]{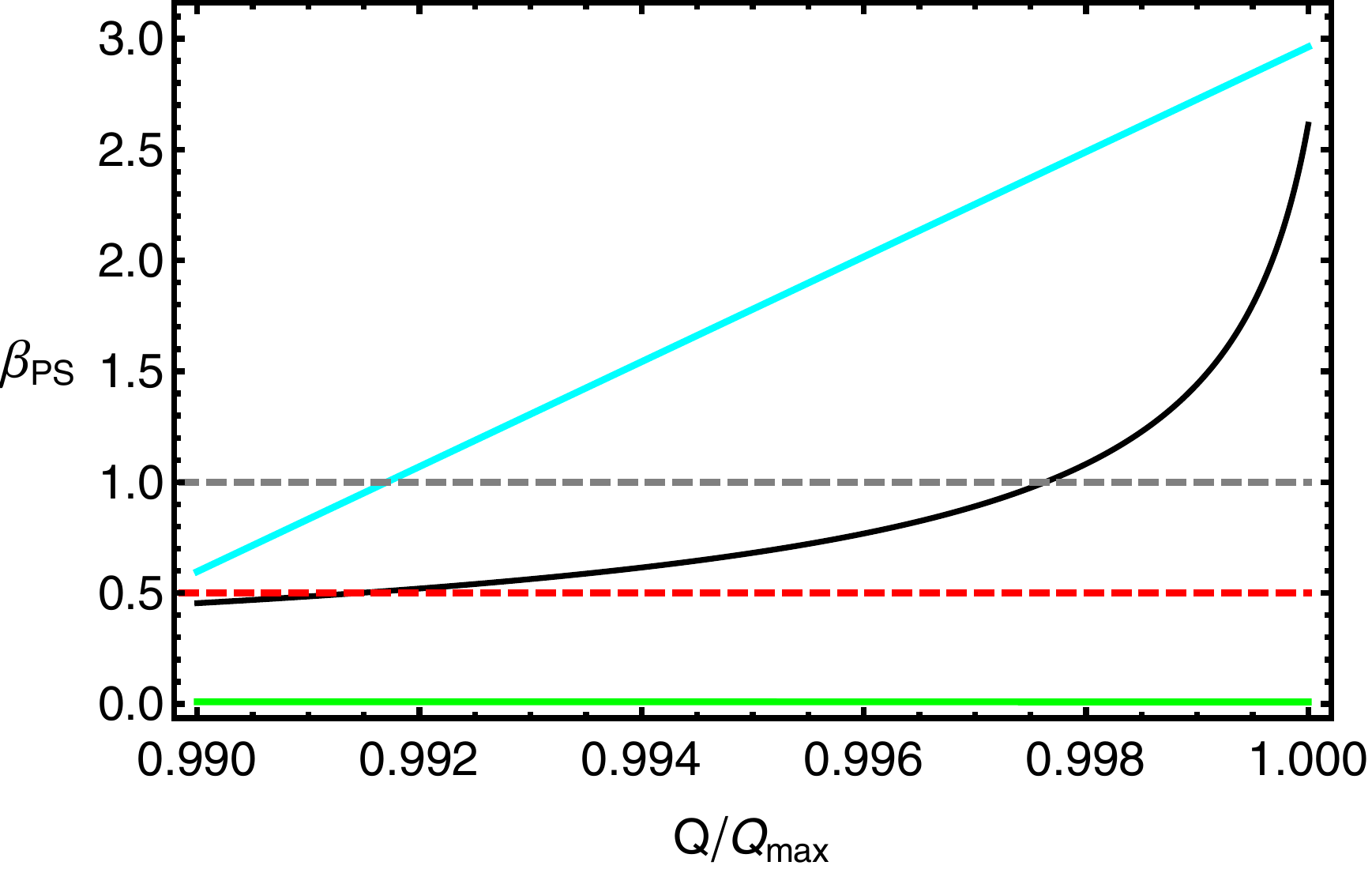}
	\endminipage\hfill
	\minipage{0.32\textwidth}%
	\includegraphics[width=\linewidth]{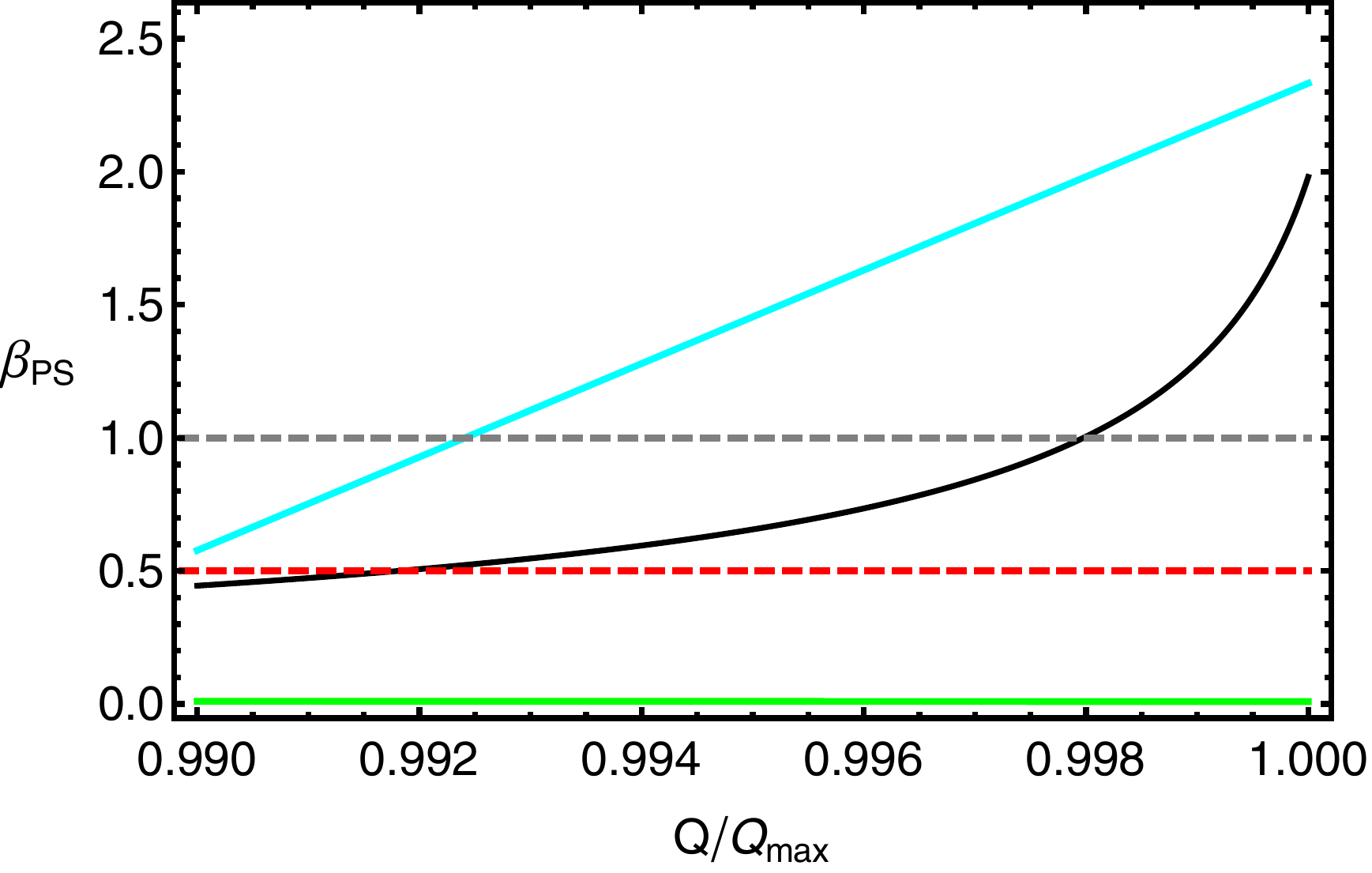}
	\endminipage\hfill
	\caption{  The variation of $\beta$ arising from PS-QNMs in terms of electric charge $Q$ for different values of 	$\eta:-2\times10^{-2},0,2\times10^{-2}$ respectively correspond to curves: cyan, black, and green with fixed values $\Lambda: 10^{-3}, 3\times10^{-3}, 5\times10^{-3}$ from left to right. }
	\label{Modett}
\end{figure}

\begin{figure}
	\minipage{0.5\textwidth}
	\includegraphics[width=\linewidth]{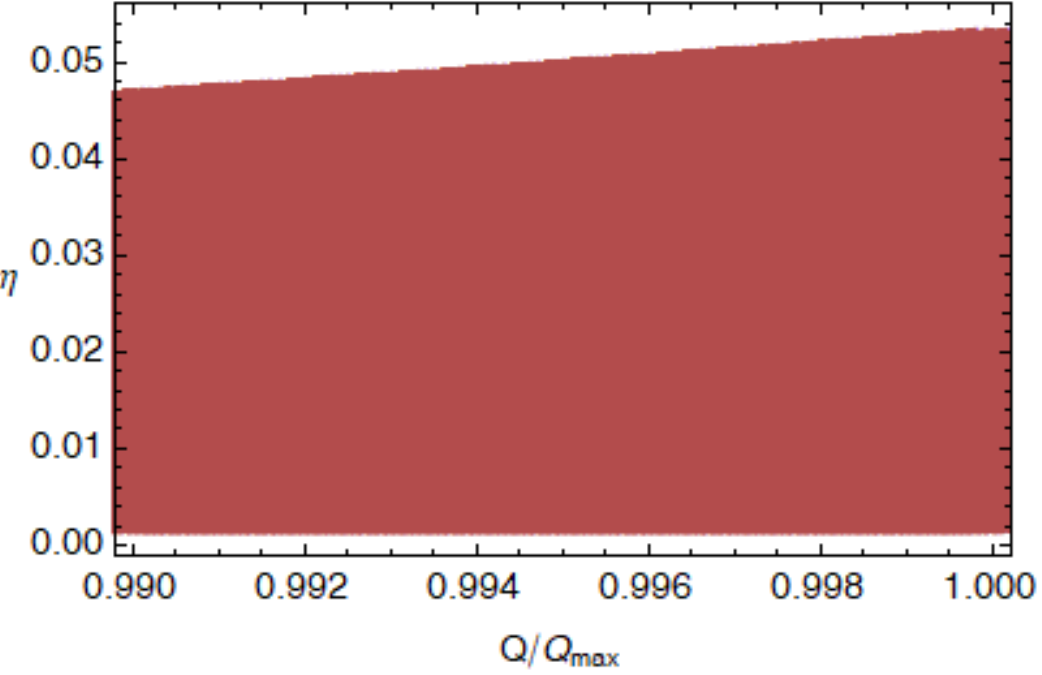}
	\endminipage\hfill
	\minipage{0.5\textwidth}
	\includegraphics[width=\linewidth]{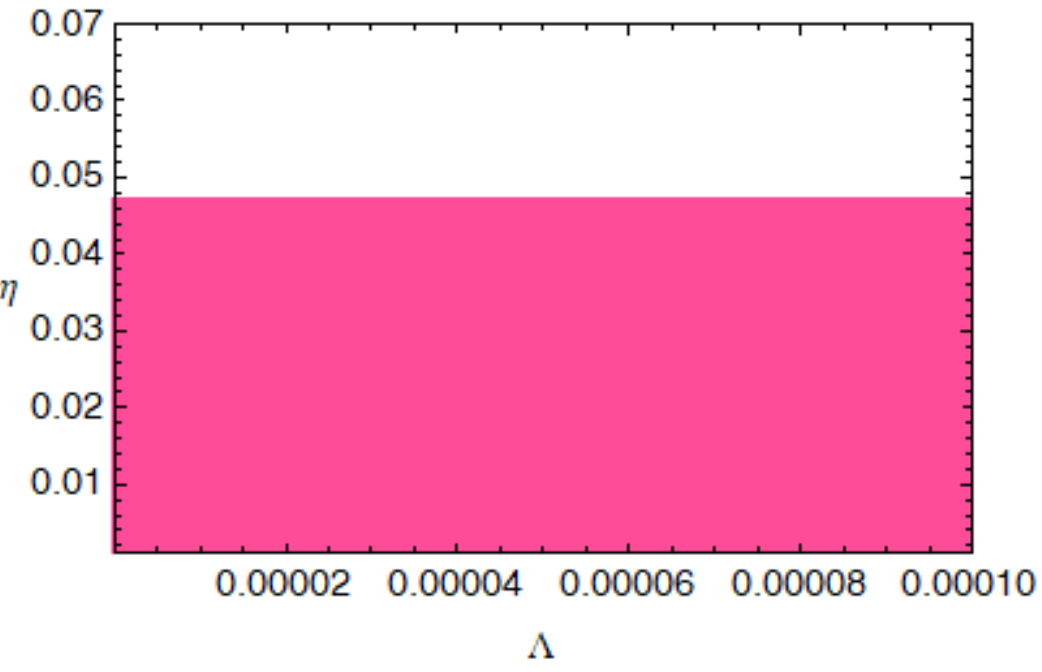}
	\endminipage\hfill
	\caption{ The colored-shaded area in the left and right panels are respectively the allowed region
		of $Q/Q_{max}-\eta$  and $\Lambda-\eta$ parameter spaces in which the $\beta$-criterion
		related to PS-QNMs is placed within $0<\beta_{PS}<1/2$. The colorless region is actually the exclusion region. Note that in the right panel low limit of $\eta$ and $\Lambda$ is very small but not necessarily zero.}
	\label{Scan2}
\end{figure}

Now we proceed to the analysis of the extracting of the $\beta$-criterion of PS-QNMs.
From Eqs. (\ref{rphnospin}) and (\ref{SG0}) we have
\begin{equation}\label{rpEM}
	\begin{aligned}
		&5 r^4 \left(r (r-3 M)+2 Q^2\right)-8 \eta  Q^4=0~, \\
		&5 r^5 \left(6 M+\Lambda  r^3-3 r\right)-15 Q^2 r^4+6 \eta  Q^4=0~,
	\end{aligned}
\end{equation} where respective result in releasing the location of PCO ($r_{\rm ph}$) and relevant horizons ($r_{\rm c,e}$).  By putting the metric solution (\ref{SG0}) into (\ref{lyanospin}), (\ref{betastatic}), and taking on account $r_{\rm ph}$ and $r_{\rm c}$ from solving (\ref{rpEM}) then for the Lyapunov exponent $\lambda _{\rm EMSG}$ and deterministic factor $\beta^{\rm EMSG}_{\rm PS}$, we have
\begin{equation}\label{lSM}
	\lambda _{\rm EMSG}=\frac{\sqrt{\left(2 Q^2 r_{\text{ph}}^4-r_{\text{ph}}^6-8 \eta  Q^4\right) \left(5 r_{\text{ph}}^4 \left(6 M r_{\text{ph}}+\Lambda  r_{\text{ph}}^4-3
			r_{\text{ph}}^2-3 Q^2\right)+6 \eta  Q^4\right)}}{\sqrt{15} r_{\text{ph}}^7}~,
\end{equation}
and
\begin{equation}
	\begin{aligned}
		\beta^{\rm EMSG}_{\rm PS}=\frac{\sqrt{\left(-\frac{4 M}{r_{\text{ph}}}-\frac{4 \eta  Q^4}{5 r_{\text{ph}}^6}+\frac{2Q^2}{r_{\text{ph}}^2}-\frac{2\Lambda  r_{\text{ph}}^2}{3}+2\right) b}}{\left(\frac{4 M}{r_c^2}+\frac{24 \eta  Q^4}{5
				r_c^7}-\frac{4 Q^2}{r_c^3}-\frac{4 \Lambda  r_c}{3}\right)}~, \\
	\end{aligned}
\end{equation}
where
\begin{equation}\label{b}
	\begin{aligned}
		b=\left(-\frac{2 M}{r_c^2}-\frac{12 \eta  Q^4}{5 r_c^7}+\frac{2 Q^2}{r_c^3}+\frac{2 \Lambda  r_c}{3}\right) \left(-\frac{2 \Lambda }{3}-\frac{4
			M}{r_{\text{ph}}^3}-\frac{84 \eta  Q^4}{5 r_{\text{ph}}^8}+\frac{6 Q^2}{r_{\text{ph}}^4}\right)-\frac{2 \Lambda }{3}-\frac{4 M}{r_{\text{ph}}^3}-\frac{4
			\eta  Q^4}{5 r_{\text{ph}}^8}+\frac{2 Q^2}{r_{\text{ph}}^4}+\frac{2}{r_{\text{ph}}^2}~.
	\end{aligned}
\end{equation}

In Fig. (\ref{Modett}) plotted the dimensionless deterministic factor $\beta_{PS}$  against the dimensionless parameter $Q/Q_{max}$, for different choices of $\eta$ and the cosmological constant $\Lambda$. As is evident, for the positive free parameter $\eta>0$ there is the possibility of satisfying the condition $0<\beta_{PS}<1/2$.
In this regard, we also perform a scan of the underlying parameters in  Fig. (\ref{Scan2}) which reveals some allowed regions in which $0<\beta_{PS}<1/2$.
In other words, by limiting the value of model parameter $\eta$ within some given range, it is then possible to control the rate of decay of the scalar field perturbation outside the black hole to keep the SCCC. Similar to the $f(R)$ gravity case, here also we see from the right panel of Fig. (\ref{Scan2}) that by taking the model parameter $\eta>0$, one no longer has to worry about the value of the cosmological constant. Since within a given range of $\eta>0$, by setting any optional value for $\Lambda$, we are deal with unstable $\mathcal{CH}$.
It is noteworthy that this window opened to rescue the SCCC is indeed a direct result of the enrichment of the charged-dS black hole solution by EMSG-Maxwell-theory.
Something that is missing in the standard Einstein-Maxwell-theory.
As a result, it seems that by taking the model parameter $\eta$ limited to some positive values within a charged spherically symmetry background occupied by a small cosmological constant, one can hope to revive the SCCC. It also can be interpreted as a theoretical constraint on the model parameter $\eta$.

\section{Conclusion}\label{con}

General relativity as a well-behaved theory even in the presence of the Cauchy horizon able to formulate the dynamic of gravity in a deterministic manner because the strong cosmic censorship conjecture (SCCC) proposed by Penrose prohibits the extension of spacetime metric across the Cauchy horizon. However, it is shown that in the de Sitter universe with the positive cosmological constant, this conjecture is no longer valid, indicating the classical fate of an observer is not under control by the initial data. 
Namely, in a de Sitter universe governed by Einstein's equations then the deterministic feature of general relativity may not be supported by solutions such as charged and rotating black holes in the extremal limit. 
At the same time, one must accept the fact that general relativity due to some shortcomings on small and large scales can not be a perfect theory and merely is a solid effective theory that must be supplemented by some corrections. As a result, there exists a tempting possibility that SCCC may be respected when these correction terms are taken into account.

Motivated by this potential possibility, by employing the linear perturbations due to
the massless scalar field propagating on the fixed background of Reissner-Nordstr\"{o}m-de Sitter solutions admitted by two modified models of gravity: $f(R)$ and Energy-Momentum Squared Gravity, we have developed
the validity study of SCCC beyond Einstein's gravity. First of all, by investigating  the tail of effective potential corresponds to these two models at long distances, 
we found that in charged spherically symmetric background occupied by a small cosmological constant the decay of scalar wave-tail at late time occurs with a rate faster than case $\Lambda=0$. As a result, the $\beta$-criterion (\ref{beta}) can be still valid in the context of extended theories at hand so that if even for one type of QNMs it becomes smaller than half, then the SCCC is respected. We have followed this purpose by focusing on the analytical and numerical investigation of photon sphere-QNMs which are dominated in the geometric optics limit i.e. eikonal approximation with large multipole $\ell\gg 1$. 

By calculating $\beta_{PS}$ for both Einstein-Maxwell-$f(R)$ and EMSG-Maxwell-theories, we found that in the case of regarding positive values for the model parameters $f(R_0)$ and $\eta$, then it is possible to meet condition $0<\beta<1/2$. Namely, within a parameter space consisting of extremal electric charge, small cosmological constant, and relevant free parameters (within a given range) SCCC is preserved. 
In total, these windows opened to rescue SCCC are, in essence, a direct consequence of the enrichment of the charged-dS black hole solution by model parameters of these two existing modified gravities.


\vspace{1cm}
{\bf Acknowledgments:}
M. Kh would like to thank Vitor Cardoso for valuable discussions and comments.\\

\end{document}